\documentclass[journal, onecolumn]{IEEEtran}
\makeatletter
\def\endthebibliography{%
  \def\@noitemerr{\@latex@warning{Empty `thebibliography' environment}}%
  \endlist
}
\makeatother
\IEEEoverridecommandlockouts
\usepackage{cite}
\usepackage{amsmath,amssymb,amsfonts}
\usepackage{cases}
\usepackage{algorithm}
\usepackage{algorithmicx}
\usepackage{algpseudocode}
\usepackage{graphicx}
\usepackage{textcomp}
\usepackage{xcolor}
\usepackage{mathrsfs}
\usepackage{float}
\usepackage{subfigure}
\usepackage{bm}
\usepackage{color}
\usepackage{diagbox}

\def\BibTeX{{\rm B\kern-.05em{\sc i\kern-.025em b}\kern-.08em
    T\kern-.1667em\lower.7ex\hbox{E}\kern-.125emX}}
\begin{document}

\title{The optimal network throughputs when the model-aware node coexists with other nodes using different MAC protocols
}

\author{\IEEEauthorblockN{Xiaowen Ye${ ^1}$, Yiding Yu${ ^2}$, and Liqun Fu${ ^1}$}\\
\IEEEauthorblockA{		
		${ ^1}$School of Informatics, and the Key Laboratory of Underwater Acoustic Communication and Marine Information Technology of Ministry of Education, Xiamen University, Xiamen 361005, China\\
        ${ ^2}$Department of Information Engineering, The Chinese University of Hong Kong, Shatin, Hong Kong Special Administrative Region, China.\\
        Email: xiaowen@stu.xmu.edu.cn, yy016@ie.cuhk.edu.hk, liqun@xmu.edu.cn}
}

\maketitle

\section{Abstract}
In this document, we give the optimal network throughput when the DR-DLMA node (see our paper for definition) coexists with the nodes using other protocol (e.g., TDMA and/or ALOHA). Then we use the optimal network throughput as the benchmark for our paper\footnote{Deep Reinforcement Learning Based MAC Protocol for Underwater Acoustic Networks}. The simulation results in our paper demonstrated that the DR-DLMA node can achieve performance similar to that of the model-aware node.

\section{Introduction}
We consider a underwater acoustic network (UWAN) consisting of $N$ nodes and a buoy as an access point (AP) in a 3D area (see our paper). All nodes in the network transmit on a shared channel. Different nodes in the network may adopt different transmission strategies. Specifically, we assume that at least one node uses our proposed DR-DLMA protocol, and the other nodes use the TDMA or ALOHA protocol. The aim of the DR-DLMA node is to make full use of the available time slots that are not used by other nodes or resulted from propagation delays, so as to maximize the throughput of the overall UWAN. In order to get the optimal network throughputs in various scenarios, we use \textbf{the model-aware node} that knows the propagation delays and the transmission strategies of other nodes to replace the DR-DLMA node.

\section{Coexistence with TDMA Networks}\label{sec:TDMA}
We first consider the coexistence of one model-aware node and one TDMA node. Let $D_1$ and $D_2$ denote the propagation delay from the model-aware node and the TDMA node to AP, respectively. Suppose that the TDMA node transmits a data packet in time slot $t$, which will reach AP in time slot $t+D_2$. If the model-aware node transmits a data packet in time slot $t+D_2-D_1$, a collision will occur at AP. In order to maximize the total network throughput without collisions, the optimal transmission policy of the model-aware node is to transmit in all time slots except time slot $t+D_2-D_1$. As a result, the optimal network throughput is $1$.


Furthermore, when multiple model-ware nodes coexist with multiple TDMA nodes, we assume that multiple model-aware nodes are aware of each other and can cooperate with each other to make their actions in each time slot. The cooperation manner among model-aware nodes is centralized. Specifically, a model-aware node in the network is designated as a gateway, which associates with all other model-aware nodes in the network and coordinates the coexistence of the model-aware network with other networks (i.e. TDMA networks). In each time slot, the gateway decides whether a model-aware node should transmit or not. If TRANSMIT, the gateway will choose one of the model-aware nodes in a round-robin fashion to transmit. As a result of the transmission, the selected model-aware node will receive a feedback/reward from the environment after a certain propagation delay, and then immediately communications with the gateway with this information. If NOT TRANSMIT, all model-aware nodes will keep silent. The coordination information from the gateway to other model-aware nodes can be sent through a control channel. For example, the control channel can be implemented as a quite short time slot after each time slot of the environmental feedback information transmission. Therefore, all model-aware nodes can cooperate to fully utilize the available time slots that are not used by TDMA nodes. Then the optimal total network throughput is also $1$.

\section{Coexistence with ALOHA Networks}
We first consider the coexistence of one model-aware node and one ALOHA node. Let $D_1$ and $D_2$ denote the propagation delay from the model-aware node and the ALOHA node to AP, respectively. For this coexistence scenario, in each time slot, the model-aware nodes determine whether to transmit or not according to the transmission probability of the ALOHA node. We assume that the ALOHA node transmits a data packet with probability $q$ in time slot $t-D_2$ and the transmission probability of the model-aware node is $b$ in time slot $t-D_1$, then the total network throughput in time slot $t$ can be calculated as follows:
\begin{align}
{f(b)}\,=\,b(1-q)+(1-b)q.
\end{align}
Thus
\begin{equation}
\begin{gathered}
\frac{\mathrm{d}f(b)}{\mathrm{d}b}=(1-q)-q=1-2q,
\end{gathered}
\end{equation}
\begin{equation}
\begin{gathered}
\frac{\mathrm{d^2}f(b)}{\mathrm{d}b^2}=0,
\end{gathered}
\end{equation}
indicating that $f(b)$ is convex in $b$. When ${\mathrm{d}f(b)}/{\mathrm{d}b}\,\textless\,0$, i.e., $q{\textgreater}0.5$, if $b=0$, $f(b)$ can get the maximum value. When ${\mathrm{d}f(b)}/{\mathrm{d}b}\,\geq\,0$, i.e., $q{\leq}0.5$, if $b=1$, $f(b)$ can get the maximum value. As a result, when the ALOHA node transmits in time slot $t-D_2$ with probability $q$, the optimal transmission policy of the model-aware node is as follows: if $q{\textgreater}0.5$, the model-aware node does not transmit in time slot $t-D_1$; otherwise, the model-aware node transmits in time slot $t-D_1$, i.e.,
\begin{numcases}{b^*=}
0,&\text{\,if $q\,\textgreater\,0.5$},\notag\\
1,&\text{\,otherwise}.
\end{numcases}\par

Since the ALOHA node has the same transmission probability in each time slot, the optimal transmission strategy for the model-aware node is as follows: if $q{\textgreater}0.5$, the model-aware node does not transmit in each time slot; otherwise, the model-aware node transmits in each time slot. Then the optimal network throughput is
\begin{numcases}{}
q,&\text{\,if $q\,\textgreater\,0.5$},\notag\\
1-q,&\text{\,otherwise}.\label{1}
\end{numcases}\par

Furthermore, we now consider the coexistence of multiple model-aware nodes and $N$ ALOHA nodes in the network. To derive the optimal result achieved by model-aware nodes, we assume that multiple model-aware nodes are aware of each other and can cooperate with each other in each time slot (the same as in Section \ref{sec:TDMA}). Specifically, the cooperation manner among model-aware nodes is centralized. A model-aware node in the network is designated as a gateway, which associates with all other model-aware nodes in the network and coordinates the coexistence of the model-aware network with the ALOHA network. We assume that the propagation delays from the ALOHA node $i$ ($i=1,2,\cdots,N$) and the designated model-aware gateway to AP are $D_i$ and $D_{N+1}$, respectively. Let $q_i$ ($i={1,2,\cdots,N}$) denote the transmission probability of ALOHA node $i$ in time slot $t-D_i$, and $b$ denote the transmission probability of the designated model-aware gateway in time slot $t-D_{N+1}$. Then the total network throughput in time slot $t$ can be calculated as follows:
\begin{align}
{f(b)}\,=\,b\prod_{{i=1}}^{N}(1-q_i)+(1-b)\sum_{i=1}^{N}\big{(}q_i\prod_{{j=1}_{j\neq{i}}}^{N}(1-q_j)\big{)},
\end{align}
Thus
\begin{equation}
\begin{gathered}
\frac{\mathrm{d}f(b)}{\mathrm{d}b}=\prod_{i=1}^{N}(1-q_i)-\sum_{i=1}^{N}\big{(}{q_i}\prod_{{j=1}_{j{\neq}i}}^{N}(1-q_j)\big{)},
\end{gathered}
\end{equation}
\begin{equation}
\begin{gathered}
\frac{\mathrm{d^2}f(b)}{\mathrm{d}b^2}=0,
\end{gathered}
\end{equation}
indicating that $f(b)$ is convex in $b$. When ${\mathrm{d}f(b)}/{\mathrm{d}b}\,\textless\,0$, if $b=0$, $f(b)$ can get the maximum value. When ${\mathrm{d}f(b)}/{\mathrm{d}b}\,\geq\,0$, if $b=1$, $f(b)$ can get the maximum value. As a result, the optimal transmission policy of the designated model-aware gateway is as follows: the designated model-aware gateway chooses to not transmit in time slot $t-D_{N+1}$ when ${\mathrm{d}f(b)}/{\mathrm{d}b}\,\textless\,0$; the designated model-aware gateway chooses one of the model-aware nodes in a round-robin fashion to transmit in time slot $t-D_{N+1}$ when ${\mathrm{d}f(b)}/{\mathrm{d}b}\,\geq\,0$, i.e.,
\begin{numcases}{b^*=}
0,&\text{\,if ${\mathrm{d}f(b)}/{\mathrm{d}b}\,\textless\,0$},\notag\\
1,&\text{\,otherwise}.
\end{numcases}\par

Since each ALOHA node has the same transmission probability in each time slot, the optimal transmission strategy for the designated model-aware gateway is as follows: if ${\mathrm{d}f(b)}/{\mathrm{d}b}\,\textless\,0$, the designated model-aware gateway chooses to not transmit in each time slot; otherwise, the designated model-aware gateway chooses one of the model-aware nodes in a round-robin fashion to transmit in each time slot. Then the optimal network throughput is
\begin{numcases}{}
\sum_{i=1}^{N}\big{(}q_i\prod_{{j=1}_{j\neq{i}}}^{N}(1-q_j)\big{)},&\text{\,if ${\mathrm{d}f(b)}/{\mathrm{d}b}\,\textless\,0$},\notag\\
\prod_{i=1}^{N}(1-q_i),&\text{\,otherwise}.\label{1}
\end{numcases}

\section{Coexistence with TDMA and ALOHA Networks}\label{subsec:qTDMA}
We first consider the coexistence of one model-aware node, one TDMA node, and $N$ ALOHA nodes. Let $D_i$ ($i=1,2,\cdots,N$), $D_{N+1}$, and $D_{N+2}$ denote the propagation delay from the ALOHA node $i$, the TDMA node, and the model-aware node to AP, respectively. Let $q_i$ denote the transmission probability of ALOHA node $i$ ($i={1,2,\cdots,N}$) in time slot $t-D_i$. Suppose that the TDMA node transmits a data packet in time slot $t-D_{N+1}$, which will reach AP in time slot $t$. If the model-aware node transmits a data packet in time slot $t-D_{N+2}$, collisions will occur at AP. Therefore, in the time slot $t-D_{N+2}$, the model-aware node refrains from transmission, then the network throughput in time slot $t$ is
\begin{align}
\prod_{i=1}^{N}(1-q_i).
\end{align}
In the time slots except time slot $t-D_{N+2}$, the model-aware node decides whether to transmit or not according to the value of $N$ and $q_i$ (i.e., the model-aware node will coexist with ALOHA nodes). Specifically, let $b$ denote the transmission probability of the model-aware node in each time slot except time slot $t-D_{N+2}$, then the network throughput in a certain time slot except time slot $t$ is
\begin{align}
b\prod_{i=1}^{N}(1-q_i)+(1-b)\sum_{i=1}^{N}\big{(}q_i\prod_{{j=1}_{j\neq{i}}}^{N}(1-q_j)\big{)}.
\end{align}
Let $p$ denote the ratio of the number of time slots used by the TDMA node in a frame to the total number of time slots in a frame, then the average network throughput in each time slot can be calculated as follows:
\begin{align}
{F(b)}\,=\,p\prod_{i=1}^{N}(1-q_i)+(1-p){\Big{(}}b\prod_{i=1}^{N}(1-q_i)+(1-b)\sum_{i=1}^{N}\big{(}q_i\prod_{{j=1}_{j\neq{i}}}^{N}(1-q_j)\big{)}{\Big{)}}.
\end{align}
Thus,
\begin{align}
\frac{\mathrm{d}F(b)}{\mathrm{d}b}=&(1-p){\Big{(}}\prod_{i=1}^{N}(1-q_i)-\sum_{i=1}^{N}\big{(}q_i\prod_{{j=1}_{j\neq{i}}}^{N}(1-q_j)\big{)}{\Big{)}},
\end{align}
\begin{equation}
\begin{gathered}
\frac{\mathrm{d^2}F(b)}{\mathrm{d}b^2}=0,
\end{gathered}
\end{equation}
indicating that $F(b)$ is convex in $b$. When ${\mathrm{d}F(b)}/{\mathrm{d}b}\,\textless\,0$, if $b=0$, $F(b)$ has the maximum value:
\begin{align}
p\prod_{i=1}^{N}(1-q_i)+(1-p)\sum_{i=1}^{N}\big{(}q_i\prod_{{j=1}_{j\neq{i}}}^{N}(1-q_j)\big{)}.
\end{align}
When ${\mathrm{d}F(b)}/{\mathrm{d}b}\,\geq\,0$, if $b=1$, $F(b)$ has the maximum value:
\begin{align}
\prod_{i=1}^{N}(1-q_i).
\end{align}\par
As a result, if the TDMA node transmits in time slot $t-D_{N+1}$, then the optimal transmission policy of the model-aware node is as follows: in time slot $t-D_{N+2}$, the model-aware node refrains from transmission. In the time slots except time slot $t-D_{N+2}$, if ${\mathrm{d}F(b)}/{\mathrm{d}b}{\geq}0$, the model-aware node transmits; otherwise, it does not transmit. Let $z={\mathrm{d}F(b)}/{\mathrm{d}b}$, then the optimal network throughput can be summarized as follows:
\begin{numcases}{}
p\prod_{i=1}^{N}(1-q_i)+(1-p)\sum_{i=1}^{N}\big{(}q_i\prod_{{j=1}_{j\neq{i}}}^{N}(1-q_j)\big{)},&\text{$z\textless0$},\notag\\
\prod_{i=1}^{N}(1-q_i),&\text{$z\geq0$}.\label{2}
\end{numcases}\par

Furthermore, when multiple model-aware nodes coexist with $M$ TDMA nodes and $N$ ALOHA nodes, we assume that model-aware nodes are aware of each other and can cooperate with each other (the same as in Section \ref{sec:TDMA}). Specifically, the cooperation manner among model-aware nodes is centralized. A model-aware node in the network is designated as a gateway, which associates with all other model-aware nodes in the network and coordinates the coexistence of the model-aware network with other networks (i.e., TDMA networks and ALOHA networks). Let $D_i$, $D_j$, and $D_{N+M+1}$ denote the propagation delays from ALOHA node $i$ ($i={1,2,\cdots,N}$), TDMA node $j$ ($j={N+1,N+2,\cdots,N+M}$), and the designated model-aware gateway, respectively. Let $q_i$ denote the transmission probability of ALOHA node $i$ in time slot $t-D_i$. Suppose that TDMA node $j$ transmits a data packet in time slot $t-D_j$, then the optimal transmission strategy of the designated model-aware gateway is as follows: in the time slot $t-D_{N+M+1}$, the designated model-aware gateway refrains from transmission. In the time slots except time slot $t-D_{N+M+1}$, if $z\textless0$, the designated model-aware gateway chooses to not transmit; otherwise, the designated model-aware gateway chooses one of the model-aware nodes in a round-robin fashion to transmit. Let $p$ denote the total ratio of the number of time slots used by TDMA nodes in a frame to the total number of time slots in a frame, then the optimal network throughput is the same as \eqref{2}.

\bibliographystyle{IEEEtran}
\vspace{12pt}

\end{document}